\documentclass[aps,prb,twocolumn]{revtex4-1}

\usepackage{graphicx}

\begin{document}

\title{Atomistic Mechanisms of Nonlinear Graphene Growth on Ir Surface}

\author{Ping Wu}
\author{Huijun Jiang}
\author{Wenhua Zhang}
\author{Zhenyu Li}
\thanks{Corresponding author. E-mail: zyli@ustc.edu.cn}
\author{Zhonghuai Hou}
\thanks{Corresponding author. E-mail: hzhlj@ustc.edu.cn}
\author{Jinlong Yang}

\affiliation{Hefei National Laboratory for Physical Sciences at
     Microscale, University of Science and Technology of
     China, Hefei, Anhui 230026, China}

\begin{abstract}
As a two-dimensional material, graphene can be naturally obtained via epitaxial growth on a suitable substrate. Growth condition optimization usually requires an atomistic level understanding of the growth mechanism. In this article, we perform a mechanistic study about graphene growth on Ir(111) surface by combining first principles calculations and kinetic Monte Carlo (kMC) simulations. Small carbon clusters on the Ir surface are checked first. On terraces, arching chain configurations are favorable in energy and they are also of relatively high mobilities. At steps, some magic two-dimensional compact structures are identified, which show clear relevance to the nucleation process. Attachment of carbon species to a graphene edge is then studied. Due to the effect of substrate, at some edge sites, atomic carbon attachment becomes thermodynamically unfavorable. Graphene growth at these difficult sites has to proceed via cluster attachment, which is the growth rate determining step. Based on such an inhomogeneous growth picture, kMC simulations are made possible by successfully separating different timescales, and they well reproduce the experimentally observed nonlinear kinetics. Different growth rates and nonlinear behaviors are predicted for different graphene orientations, which is consistent with available experimental results. Importantly, as a phenomenon originated from lattice mismatch, inhomogeneity revealed in this case is expected to be quite universal and it should also make important roles in many other hetero-epitaxial systems.
\end{abstract}

\maketitle

\section{introduction}
Graphene, a monolayer of $sp^2$ hybridized carbon atoms, has attracted an intense research interest recently due to its unique electronic structure and great application potential. \cite{Novoselov200466, Geim2007review, Geim2009review, Schwierz2010review} Currently, there are several ways available to produce graphene. The most elegant method using micromechanical exfoliation is ready to produce high quality samples, \cite{Novoselov200466} but it is difficult to be scaled up. Solution based mass production, on the other hand, is challenging to reach a high sample quality. \cite{Dreyer201040} Epitaxial growth can generate large graphene sample on metal surfaces potentially with high quality, which makes it the method of choice for large scale electronic applications of graphene. Various metal surfaces (including Ni, Cu, Ru, Ir, Pd, Co, Pt, and etc) have been used to grow graphene. \cite{Ru2008, Coraux200808, CuScience, Kim200906, Zeng201185} Notice that, on different substrates, the graphene growth behavior can be very different. \cite{Li200968}

Ir substrate is very attractive for graphene growth, where large-scale samples with long-range order and continuity can be obtained. Graphene can spread over step edges on Ir surface just like a carpet, \cite{Coraux200808} and it is typically well aligned with the Ir substrate, forming a moir\'{e} pattern. \cite{NDiayeNJP, Ir-Pletiko2009102, Loginova200980PRB} Interestingly, the moir\'{e} pattern can then serve as a template to grow patterned structure of metal clusters, with a great potential in nanocatalysis and nanomagnetism. \cite{NDiaye200697} At the same time, our previous study has shown that Ir surface is very rigid compared to the soft Cu surface, \cite{Wu201001, Zhang201182} which makes it a good model system to study the growth mechanism of graphene.

On Ir(111) surface, several graphene orientations have been identified . \cite{Loginova200980PRB} Compared to the majority one (R0), three other orientations are rotated by $14\,^{\circ}$ (R14), $18.5\,^{\circ}$ (R18.5), and $30\,^{\circ}$ (R30), respectively. This observation indicates that the interaction between graphene and Ir substrate is not too weak to be totally orientation independent, and it is also not too strong to only pick up a single orientation. A recent angle-resolved photoemission spectroscopy (ARPES) experiment has indicated that graphene is chemisorbed with its R0 orientation while physisorbed with the R30 orientation. \cite{Ir-Starodub} Therefore, different orientations lead to different electronic structures, which then provides a flexibility to tune the electronic structure of a grown graphene by controlling its orientation. \cite{APL-select}

%First principles calculations predict that interaction between graphene and Ir surface is weak and electronic structure of graphene is not affected by the surface, \cite{NDiaye200697,Feibelman200877} which is confirmed by later angle-resolved photoemission spectroscopy (ARPES) experiment. \cite{Ir-Pletiko2009102}

With multiple orientational variants, the growth of graphene on Ir surface is complicated. \cite{Loginova2009NewJP} Typically, the R0 phase nucleates first. Then, at the boundary of a growing R0 island, one of the three minority orientations (mainly R30) can occasionally also nucleate. Once nucleated, the R30 phase grows much faster than the R0 phase. Most interestingly, growth rate of the dominant R0 phase has a nonlinear (approximately quintic) relationship with the concentration of carbon adatoms, which indicates that five-membered cluster attachment makes an important role in the R0 phase growth. \cite{Loginova2008NewJP} A homogeneous kinetic model has been constructed accordingly by fitting experimental data for a similar surface, Ru(0001). \cite{Zangwill201192} However, the atomistic mechanism of graphene growth on Ir surface is still unclear.

%linear, quadratic, cubic, quartic, quintic

In this article, first-principles calculations are carried out to study graphene growth on Ir(111) surface. Adsorption, diffusion, and coalescence of small carbon species are considered first. Then, nucleation at step sites is discussed. Finally, the growth process is studied by attaching small carbon species to a graphene edge described with a nanoribbon model. With all these elementary processes understood, the overall picture of graphene growth on Ir surface is then obtained, which is characterized by inhomogeneous growth and orientation sensitivity. Based on this picture, a multiscale kinetic Monte Carlo (kMC) model is constructed, which well reproduces the experimentally observed nonlinear growth behavior.

\section{Computational Details}

All calculations were carried out with plane-wave based density functional theory (DFT) implementation in the Vienna ab initio simulation package (VASP). \cite{JCMS96, PRB96} A 400 eV kinetic energy cutoff was chosen for plane waves. The Perdew-Burke-Ernzerhof (PBE) functional \cite{PRL96} was used to describe exchange and correlation. K-point sampling within the Monhorst-Pack scheme \cite{PRB1989} was carefully tested to produce well converged results. A four-layer slab model was used to describe the Ir(111) surface, with the bottom layer fixed to its optimized bulk geometry (the corresponding Ir-Ir bond length is 2.74 \AA). Repeated slabs were separated by more than 10 {\AA} to avoid interaction between each other. Vicinal (322) and (332) surfaces were used to model the stepped Ir(111) surface, which contains \{100\} (A-type) and \{111\} (B-type) microfacets, respectively. Graphene growth front was modeled using graphene nanoribbons. For the three minority orientations of graphene, only the R30 phase was considered since the other two are rarer than R30 by one order of magnitude. \cite{Loginova200980PRB}

Potential energy of carbon clusters on Ir surface is defined as
\begin{equation}
E_p=(E_{C/Ir}-E_{Ir})/N_C
\end{equation}
where $E_{C/Ir}$ and $E_{Ir}$ are energies of the absorbed system and the clean Ir surface, and $N_C$ is the number of carbon atoms in the cluster. Energy of an isolated carbon atom in a big supercell is taken as an energy reference. Climbing image nudged elastic band (CI-NEB) method \cite{Henkelman200001} was used for transition state location and barrier height determination. Residual forces were within 0.02 eV/{\AA} for both geometry optimization and transition state location.

\section{Results and discussion}

\subsection{Small carbon species on terrace}

\begin{figure}[tbhp]
\begin{center}
\includegraphics[width=8cm]{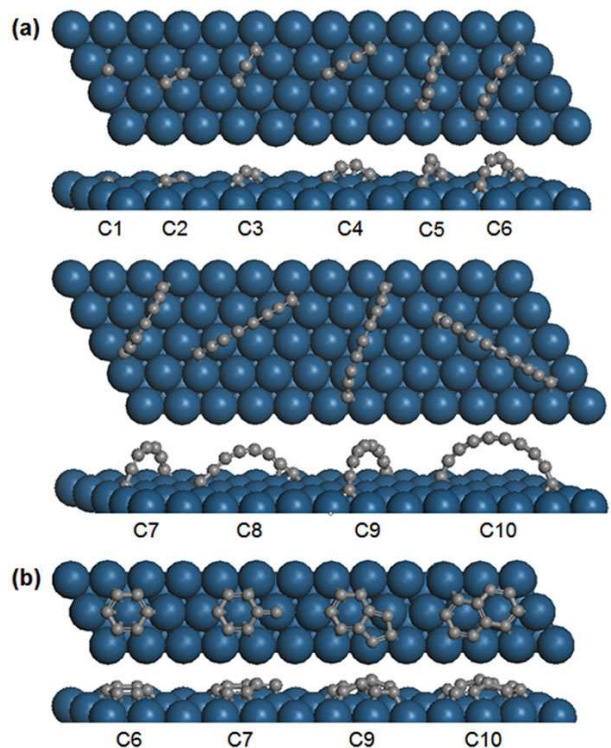}
\end{center}
\caption{(a) One dimensional chain structures of carbon clusters on Ir(111). (b) Some two dimensional compact structures.}
\label{fig:cluster}
\end{figure}

Relative stabilities of different carbon species on the surface determines their evolution trend. Here, as many as possible configurations of carbon species $C_{N}$($N$=1,\ldots,10), including both chain and compact structures, are explored. For a carbon monomer, as shown in Figure \ref{fig:cluster}a, the hexagonal close-packed (hcp) hollow site is the most favorable adsorption site ($E_p$ = -7.43 eV), which is 0.26 eV more stable than the face-centered cubic (fcc) hollow site, agreeing well with previous results. \cite{zhang} Carbon atom at a top site is 1.73 eV higher in energy than the most stable hcp hollow site, with the Ir atom under the atop carbon atom being pulled out from the surface by about 0.5 \AA. Bridge-site adsorbed carbon atom will be spontaneously relaxed to a hollow site. The subsurface octahedral site is also much more unstable (1.40 eV) than the hcp hollow site, consistent with the relatively low carbon solubility in Ir. \cite{Ir-solubility}

For small carbon clusters on the surface, generally the most stable configuration is a chain structure, with both terminal carbon atoms occupying a hollow site. Longer carbon chains prefer to form arches on the surface. Similar structures have also been predicted on Ni \cite{Ding1109} and Cu \cite{zhang11} surfaces. Another important kind of configurations are two dimensional (2D) compact structures (Figure \ref{fig:cluster}b). They are all less stable than one dimensional (1D) chain structures, except for C$_6$. Compact C$_6$ ring is slightly more stable (0.06 eV/atom) than its corresponding arching chain structure. Although clusters considered in this study are relatively small, we can already see a trend for compact structures to be dome-like, \cite{PRL-Dome} with peripheral carbon atoms bonded more strongly to the Ir surface.

\begin{figure}[tbhp]
\begin{center}
\includegraphics[width=8.5cm]{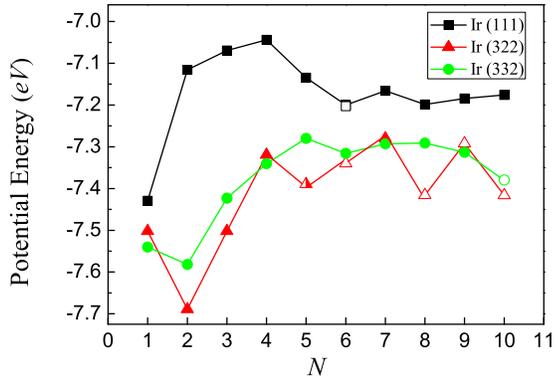}
\end{center}
\caption{Potential energy ($E_p$) of the most stable carbon species (C$_N$) on Ir(111) surface and Ir(332) and Ir(322) steps. Hollow points indicate 2D compact structures.} \label{fig:pot}
\end{figure}

Relative stabilities of carbon species with different sizes on Ir(111) terrace can be compared using the potential energy defined in Eq. (1). As a limit case, potential energy of graphene on Ir surface is interesting. Its upper bound is estimated by the binding energy of a free standing graphene monolayer (-7.99 eV). As shown in Figure \ref{fig:pot}, this value is already much lower than those of carbon species considered in this study. This is the thermodynamic driving force for graphene growth. Carbon monomer is more stable than all small clusters. For example, potential energy of carbon dimer is -7.12 eV, even higher than that of fcc-hollow carbon. C$_6$ is already more stable than fcc-hollow monomer, but still less stable than hcp-hollow carbon. Therefore, concentration of small clusters is expected to be much smaller than that of carbon monomer on Ir(111).

%The most interesting result is that, if we analysis energy according to classical nucleation theory, we get a nucleation size of 4, which is pretty small.

\begin{figure}[tbhp]
\begin{center}
\includegraphics[width=8cm]{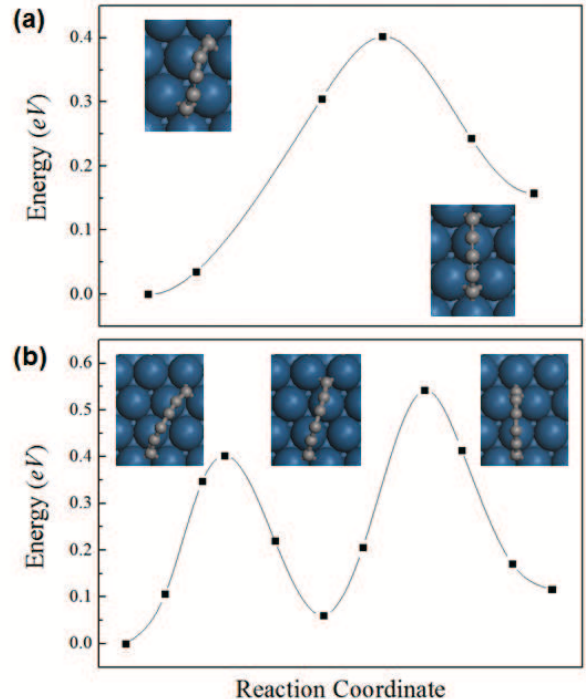}
\end{center}
\caption{Minimum energy path for (a) C$_5$ and (b) C$_6$ carbon chains diffusion on Ir(111) surface. Inset shows geometric configurations before and after each elementary diffusion step, which is characterized by hopping of one terminal carbon atom between two hollow sites (the walk-with-legs mode).} \label{fig:56}
\end{figure}

Mobility of small carbon clusters on Ir(111), which is mainly determined by the diffusion barrier, is also important for graphene growth. The diffusion barrier for a carbon monomer from hcp-hollow to fcc-hollow is 0.71 eV. As expected, mobility of C$_2$ and C$_3$ is even lower, with a diffusion barrier of 0.76 and 0.93 eV, respectively. However, for larger carbon clusters with arching chain structure, the diffusion barriers are much lower (0.40 eV for C$_5$ and 0.54 eV for C$_4$ and C$_6$). These low barriers are a result of the walk-with-legs diffusion mode of the arching chain structure (Figure \ref{fig:56}). Although not calculated, mobility of two dimensional compact configurations is expected to be much lower, since their diffusion usually requires to break several bonds simultaneously. When mobile carbon species meet, they can coalesce on the surface. The energy barrier for the combination of two neighboring carbon monomers is 1.44 eV, and it is similar (1.42 eV) for the incorporation of one carbon atom into an existing dimer. When adding a carbon atom to an arching chain structure of C$_4$, the energy barrier decreases to 0.86 eV. From the energy-barrier point of view, formation of small carbon clusters on terrace is generally more difficult than their diffusion.

\subsection{Effects of step edges}

\begin{figure}[tbhp]
\begin{center}
\includegraphics[width=8cm]{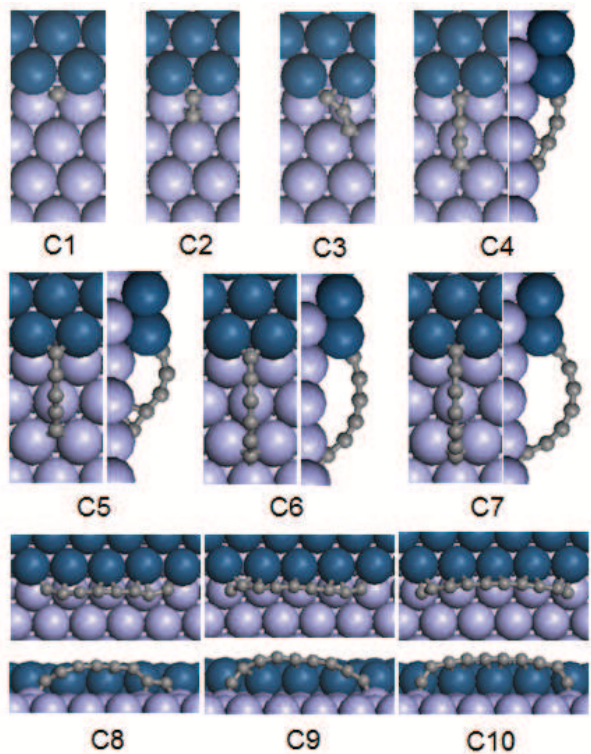}
\end{center}
\caption{Stable chain structures of carbon species on Ir(322) surface.} \label{fig:322-l}
\end{figure}

During graphene growth on Ir surface, nucleation prefers to occur at step edges. \cite{Coraux2009} It is thus very desirable to study carbon species adsorption at steps. Both A-type and B-type steps are considered by using Ir(322) and Ir(332) surfaces, respectively. By subtracting the (111) terrace contribution from surface energy, we can compare the stability of these two types of steps. The A-type step is more stable than the B-type, with their step formation energy as 0.37 and 0.45 eV/\AA, respectively. We study the more stable A-type first. The most stable adsorption site of atomic carbon is at the lower edge of the step (Figure \ref{fig:322-l}). The corresponding potential energy ($E_p$=-7.50 eV) is 0.07 eV lower than that of the hcp-hollow carbon on terrace, which means that step edge is more favorable than terrace. To escape from the step edge, a 0.86 eV barrier should be conquered. Dimer and trimer also tend to be attracted at the step edge, with a potential energy of -7.69 and -7.50 eV, respectively. The step edge fully covered with carbon monomers, however, is highly unstable ($E_p$=-7.00 eV). The reason is that carbon atoms will sink into the \{100\} microfacets, which induces a large geometry distortion. The step edge fully occupied by dimers has an $E_p$ (-7.36 eV), which is not very high and is similar to that of hcp-hollow carbon. Therefore, dimer can be trapped at the step edge with a relatively high concentration.

The most stable structure of a carbon trimer at the step edge is neither perpendicular nor parallel to the step. For larger clusters, a handle-like chain structure perpendicular to the step is energetically very favorable. In such a structure, one terminal atom occupies a bridge-like step site just as in the dimer case, while the other terminal atom occupy a hollow site on terrace. Due to the limitation of the (322) model used for the step, we can not study perpendicular carbon chain longer than seven. In those cases, we only consider carbon chains along the step edge. With significant lattice mismatch between carbon and iridium, such chains are typically not well accommodated. Structure of chains with similar length can be notably different. This also makes chains along the step edge energetically not very favorable.

\begin{figure}[tbhp]
\begin{center}
\includegraphics[width=8cm]{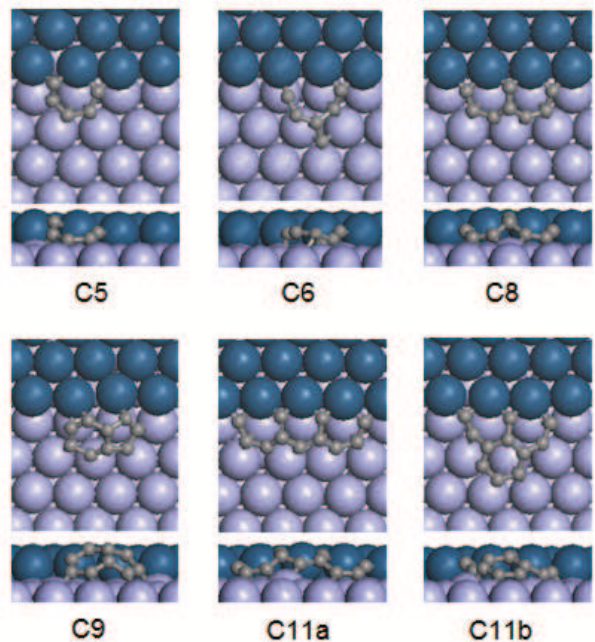}
\end{center}
\caption{Stable compact structures of carbon species on Ir(322) surface.} \label{fig:322-r}
\end{figure}

Compact structures are generally less stable than the chain structures for small carbon clusters at the step edge. However, a class of magic compact structures with high stability have been identified. The smallest example is a C$_5$ cluster (Figure \ref{fig:322-r}), which has the same energy as its corresponding handle-like chain structure ($E_p$=-7.39 eV). It can be considered as a structure relaxed from two dimers at step edge connected by another carbon atom on terrace. Since the step edge can trap many dimers, such a structure motif may act as a starting point of graphene growth. To check this possibility, we extend this structure to two rings (C$_8$). This structure is already more stable than its corresponding handle-like chain structure. More importantly, a fragment of zigzag edge with the R0 orientation appears on terrace in this structure. This is consistent with the experimental observations than R0 phase nucleates at step edges, and it prefers to grow with zigzag edges. \cite{asusc2005, Coraux2009}

Similarly, this kind of magic structures can be further extended to three rings (C$_{11}$a), which is also the most stable structure for clusters with the same size. At the same time, instead of extending along the zigzag direction, three-ring structure can also be obtained by grow an additional hexagon in front of C$_8$ (C$_{11}$b in Figure \ref{fig:322-r}). This C$_{11}$b structure can be considered as an initialization of graphene growth from a C$_8$ structure, and the intermediate between C$_8$ and C$_{11}$b is the C$_{10}$ structure. C$_{10}$, C$_{11}$a and C$_{11}$b has almost the same potential energy, similar to that of hcp-hollow carbon on terrace.

Although it is an elementary unit of graphene, hexagon is not stable at the step edge ($E_p$=-7.08 eV) compared with the handle-like chain structure of C$_6$. It is even less stable than hexagon at terrace. Therefore, graphene growth is more likely start from the semicircle C$_5$ structure instead of forming hexagon at step edges. The most stable structure of C$_6$ at step looks like half of the C$_{10}$ structure, which is 0.03 eV per carbon atom more stable than the handle-like chain structure. A compact 5-6 ring (C$_9$) becomes more stable than the corresponding chain structure along the step edge, but it is still notably less stable than neighboring C$_8$ and C$_{10}$.

\begin{figure}[tbhp]
\begin{center}
\includegraphics[width=8cm]{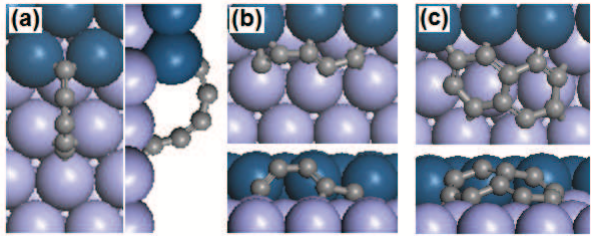}
\end{center}
\caption{(a) Handle-like structure of C$_5$ perpendicular to the Ir(332) step. (b) Chain structure of C$_5$ parallel to the Ir(332) step. (c) Compact structure of C$_{10}$ at the Ir(332) step.} \label{fig:332}
\end{figure}

Carbon clusters on the Ir(332) surface are similar to those on Ir(322) except the missing of magic compact structures. Monomer and dimer are very stable at the step edges, with a $E_p$ of -7.54 and -7.58 eV, respectively. The diffusion barrier of a carbon monomer from a step to the lower terrace is 1.05 eV. When atomic carbon fully covers the step, its potential energy becomes -7.37 eV.  Dimer fully covered step gives a potential energy still as low as -7.52 eV. At the B-type step edge, trimer can form a straight chain perpendicular to the step, while C$_4$ becomes bent. The most stable structure of C$_6$ and C$_7$ is also handle-like chain. However, for C$_5$, the handle-like structure has a too large curvature (Figure \ref{fig:332}a). Therefore, its most stable structure becomes the chain along the step edge, which is 0.02 eV lower in potential energy. Compact structure becomes more stable than chain structure until the cluster size equal to 10, where a fused two-hexagon structure is energetically favorable ($E_p$=-7.38 eV). Notice that in this structure there is also a fragment of R0 oriented zigzag edge.

\subsection{Carbon attachment at graphene edge}

Growth generally proceeds by attaching small carbon species on the edges of a growing island. Here, we use graphene nanoribbon as a model edge system. To keep the supercell reasonably small, C-C bonds are slightly stretched (within several percent) to match the Ir lattice. Consistent with experimentally studied graphene islands, which are dome-like with strong edge-substrate binding, \cite{PRL-Dome} nanoribbons also form arched structure in the width direction. To study attachment thermodynamics, we define an averaged formation energy ($E_f$) for a specific attachment process as
\begin{equation}
E_f=(E_{C+R/Ir}-E_{R/Ir}-E_{C/Ir}+E_{Ir})/N_C
\end{equation}
where $E_{C+R/Ir}$ is the total energy of an absorbed ribbon with a carbon species (size $N_C$) attached, $E_{R/Ir}$ is the energy of the absorbed ribbon before attachment, and $E_{C/Ir}$ is the energy of the carbon species absorbed on the surface far away before attachment.

\begin{figure}[tbhp]
\begin{center}
\includegraphics[width=8cm]{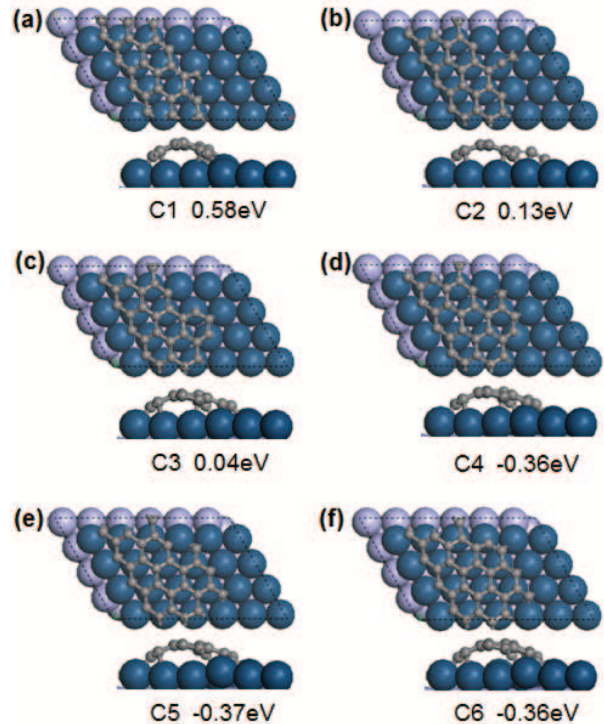}
\end{center}
\caption{Optimized structure of a zigzag edge with different carbon clusters attached. $E_f$ is marked in eV.} \label{fig:R0z}
\end{figure}

For the R0-phase, we focus on the zigzag edge, since it is preferred as observed in scanning tunneling microscopy and current imaging tunneling spectroscopy experiments. \cite{asusc2005, Coraux2009} Attaching a carbon monomer to a zigzag edge (Figure \ref{fig:R0z}) is energetically unfavorable with an $E_f$ equal to 0.58 eV. Since isolated adatom is more stable, the attached carbon atom will diffuse away much faster than its attachment. Therefore, carbon monomer attachment will not contribute to graphene growth at such a zigzag edge. In contrary to atomic carbon, four-, five-, and six-membered carbon clusters have a driving force (about 0.36 eV) to attach to the zigzag edge. With endothermic monomer attachment and exothermic cluster attachment, the experimentally observed nonlinear growth seems having a simple thermodynamic mechanism.

\begin{figure}[tbhp]
\begin{center}
\includegraphics[width=8cm]{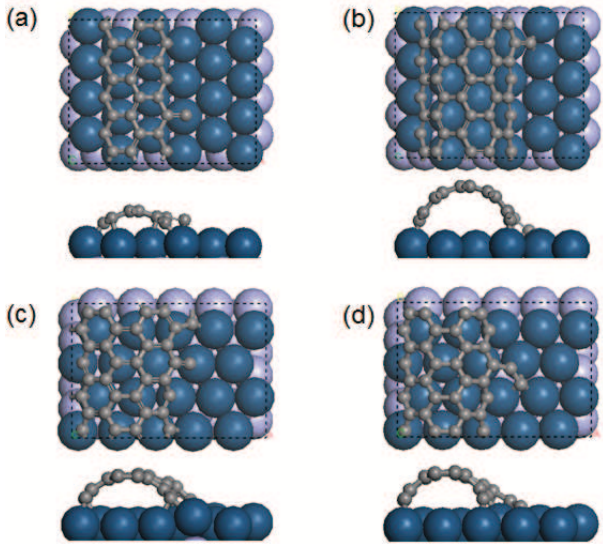}
\end{center}
\caption{(a) Optimized structure of a carbon monomer attached at zigzag R0 ribbon occupying a (a) top and (b) hollow site. (c) Two carbon monomer and (d) a carbon dimer attached R30 zigzag edge on Ir(111) surface.} \label{fig:R0}
\end{figure}

However, we also note that graphene grown on Ir(111) surface will form a moir\'{e} structure due to lattice mismatch, \cite{Loginova200980PRB,NDiaye200697} which means that different parts of graphene has different registries with the substrate lattice. For the R0 phase, in the atop-type region, all carbon atoms occupy hollow sites. However, in both hcp- and fcc-type regions, hollow site carbon is immediately neighboring to top site carbon atoms. \cite{NDiaye200697} The model presented in Figure \ref{fig:R0z} thus mainly describes edge fragment in the hcp- and fcc-type regions, where attached carbon atoms occupy top sites. Since top site is energetically very unfavorable, attachment of small clusters ($N<4$ in this case) becomes endothermic in these regions. Notice that, for monomer attachment, the attached atom is not exactly at a top site in the ribbon model adopted here. The structure with an attached atop atom (Figure \ref{fig:R0}a) is a metastable state, which leads to an $E_f$ as high as 1.98 eV. In atop-type region, the attached carbon atom is expected to occupy a hollow site instead of the top site. To describe such a behavior, we construct another ribbon model. As expected, when the attached carbon atom sits at a hollow site (Figure \ref{fig:R0}b), a different thermodynamics is obtained. The atomic carbon attachment becomes exothermic with the $E_f$ equal to -0.86 eV.

Therefore, attachment of atomic carbon can be energetically both favorable or unfavorable depending on the growth front configuration. On the kinetic side, the energy barrier for carbon monomer attachment for the models in Figures \ref{fig:R0z}a and \ref{fig:R0}b are 1.0 and 0.75 eV, respectively. These barriers are easy to be conquered at the experimental temperature. Therefore, as the most abundant species on terrace (refer to Figure \ref{fig:pot}), carbon monomer's attachment to graphene edge is expected to be a process with relatively high possibilities everywhere. However, for those thermodynamically unfavorable sites, its detachment will be much faster. At those sites, graphene growth relies on attachment of much rarer carbon clusters. Those parts thus grow much slower, and cluster attachment becomes the rate determining step of graphene growth.

\begin{figure}[tbhp]
\begin{center}
\includegraphics[width=6cm]{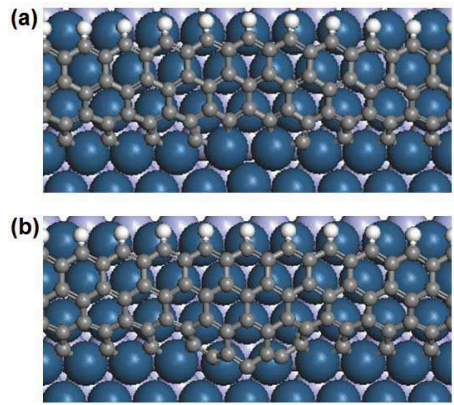}
\end{center}
\caption{(a) graphene ribbon with eight of the ten sites for carbon monomer attachment filled and (b) with an additional C$_5$ cluster attached.} \label{fig:all}
\end{figure}

To further confirm the above picture, we also perform a calculation directly using a large supercell with the period in the edge direction matching the exact moir\'{e} unit cell. \cite{NDiaye200697}  In the direction perpendicular to the edge, a narrow ribbon is used, with the other side saturated by hydrogen atoms. During geometry optimization, the $z$ coordinate of hydrogen atoms is fixed based on the experimental mean height of graphene on Ir surface. \cite{Carsten2011} Along the zigzag edge, eight of the ten sites for carbon monomer attachment is energetically favorable. Therefore, a structure with eight carbon atoms attached and a vacancy at the two top sites (Figure \ref{fig:all}a) can be easily formed, which is more stable (0.17 eV per attached carbon atom) than the unattached ribbon plus eight hcp-hollow carbon atoms. Although the vacancy is very difficult for atomic carbon attachment ($E_f$=2.11 eV), it is well fitted by a C$_5$ cluster (Figure \ref{fig:all}b). Such a C$_5$ attachment process has an $E_f$ as large as -0.70 eV. This result gives a direct evidence that some parts of the graphene edge are difficult for atomic carbon attachment and they only grow with carbon clusters supplied.

%Armchair edge is not studied in details here, because  Our carbon atom attachment thermodynamics also supports the preference of zigzag edge. For example, in the ribbon model shown in Figure \ref{fig:R0}b, even if an armchair edge is already formed, attaching an additional carbon atom at a hollow site can easily generate a zigzag edge fragment. However, it is difficult to restore the armchair edge by attaching another carbon atom, since it is on a top site (Fig. \ref{fig:R0}c).

\subsection{Kinetic Monte Carlo simulation}

Based on the physical picture from first principles, we develop a kMC model to directly reproduce the experimental nonlinear growth behavior. As already pointed out by Zangwill and Vvedensky, \cite{Zangwill201192} a brute-force kMC simulation will be impractical for graphene growth on Ir(111). There are mainly two difficulties. First, the growth front keeps moving, which requires a huge unit cell to get a statistically reliable growth rate.  Another problem is the disparate rates of different kMC events, due to the large density differences for different surface species. For example, the gap between densities of carbon monomer and arching chain $C_5$ is about $10^{15}$, which stretches the limit of even the most efficient algorithms for a direct simulation. To overcome these difficulties, we build a multiscale 'growth-front-focused' kMC model.

\begin{figure}[tbhp]
\begin{center}
\includegraphics[width=8cm]{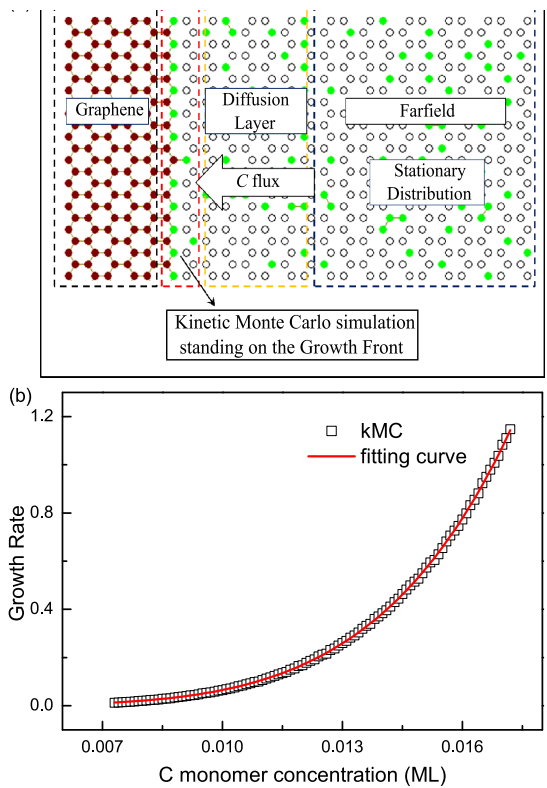}
\end{center}
\caption{(a) Schematic diagram of the kMC model standing on the graphene growth front. (b) Graphene growth rates dependence on $C$ monomer concentration obtained by kMC simulations. The solid red line is a fitted curve, and the resulting exponent is 5.17.} \label{fig:kmc}
\end{figure}

Our kMC model is based on a honeycomb lattice. The whole surface is divided into four regions: graphene region, growth front, diffusion layer, and far field (Figure \ref{fig:kmc}a). In the far field, we have an equilibrium of different $N$-sized carbon species. Their density can be obtained by considering the $N$C$_1\rightleftharpoons$ C$_N$ equilibrium \cite{Kashchiev00} with carbon potential energies calculated from first principles. Combined with first-principles diffusion barriers, we can further get the flux of difference carbon species across the diffusion layer to the graphene growth front. Then, we can focus on the carbon species attachment and detachment at the growth front, and other processes are compacted into effective carbon species fluxes. By doing so, we not only reduce the list of kMC events, but also extend the grown graphene as long as we need without extra cost.

Rates for the attachment and detachment of different carbon species can be calculated, if kinetic data of all these events are available from first principles. However, due to the effect of the Ir substrate, different sites in the honeycomb kMC lattice are not equivalent, which leads to a huge number of events. To build a tractable model, we make a simplification by considering only two kinds of sites. Based on the moir\'{e} pattern, one kind of sites can be easily occupied by carbon monomers, while the other is difficult for monomers (see Supporting Information). With such a simplification, all required kinetic data is calculated or estimated directly from first principles.

A kMC simulation with the efficient 'growth-front-focused' model described above is still prohibited by the huge gaps between fluxes of different carbon species. To solve this problem, we design a multiscale kMC algorithm similar to the nested stochastic simulation algorithm for chemical kinetic systems with disparate rates.\cite{JCP05194107} In our algorithm, the kMC events are grouped by the fluxes of the relevant carbon species. Events in different group represent the evolvement of the graphene front at different time scales. For example, attachment of C monomers represents the fastest time scale. The multiscale kMC simulation is run in the following way:

 (1) run a standard 'event-list' kMC algorithm with all events included.

 (2) once the graphene front gets stuck, which indicates that the events on the present time scales don't contribute to the graphene growth anymore, turn the simulation to the next slower time scale by turning off the events in the present time scale group.

 (3) the simulation time scale keeps jumping until the graphene grows again.

 (4) once the front of graphene moves, return to step 1.

With such a specially designed kMC algorithm, we run simulations for different $C$ monomer concentrations. Convergence of growth rate is carefully tested. The obtained relation between growth rate ($r$) and carbon monomer concentration ($m$) is fitted by $r=am^b+c$. As shown in Figure \ref{fig:kmc}b, we get a $b$ value of 5.17, which agree with the experimental value very well.  We emphasize that in our kMC simulation, all parameters are calculated or estimated from atomic scale, without any fitting to macroscale experimental data. Therefore, the good agreement between experiment and kMC simulation suggests that the growth model revealed by our first principles calculations has grasped the main physics for graphene growth on Ir(111).

\subsection{Effects of the Graphene Orientation}

Occasionally, a rotated R30 phase can be nucleated at the edge of a growing R0 phase. Based on the physical picture of the R0 phase growth, we can make a prediction about the growth kinetics of the R30 phase. We start from the moir\'{e} pattern of R30. It is different from that of the R0 phase, with more homogeneously distributed top sites. \cite{Loginova200980PRB} In the R30 phase, two top sites are separated by at least two hollow or bridge sites. We have shown that growth of the R0 phase rely on large carbon clusters to find their right place where top sites gather. Although the moir\'{e} pattern is not exactly followed at graphene edges, the more homogeneous distribution of top sites in the R30 phase is expected to remove this stringent large-cluster requirement. This will speed up the R30 phase growth compared to the R0 phase, as exactly observed in experiment.\cite{Loginova2009NewJP} It is worth to mention that understanding the different growth behaviors for different graphene orientations will enable us to control the graphene sample quality.\cite{APL-select}

\begin{figure}[tbhp]
\begin{center}
\includegraphics[width=8cm]{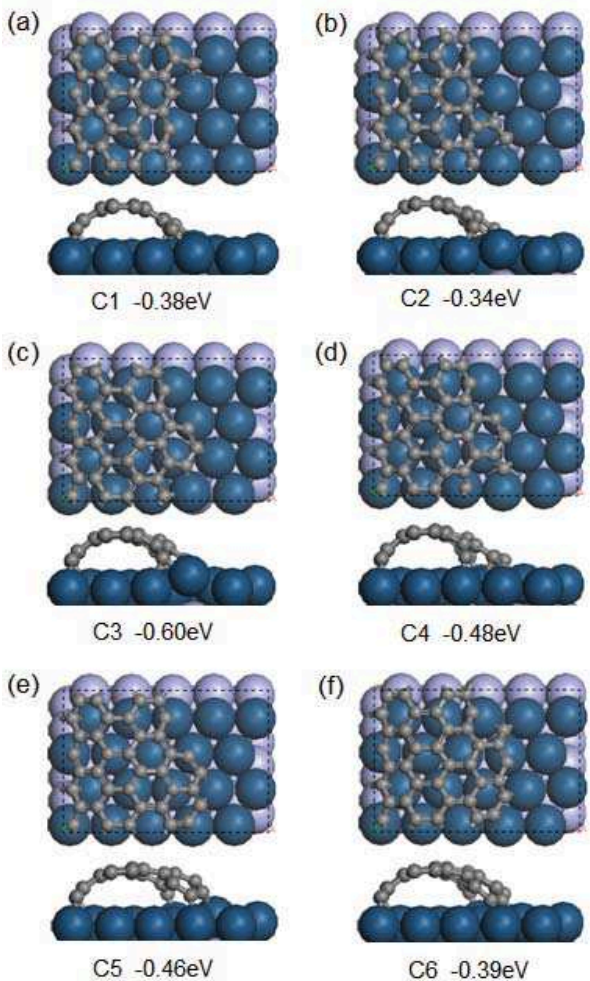}
\end{center}
\caption{Optimized structure of the R30 zigzag edge with different carbon species attached on the Ir(111) surface.} \label{fig:R30}
\end{figure}

To further study growth of the R30 phase, ribbon models are also constructed to describe the growth front. As shown in Figure \ref{fig:R30}, when a zigzag edge is chosen, carbon monomer attachment can be favorable in energy with a negative $E_f$. However, unlike the zigzag edge in the R0 phase, different edge atoms are not equivalent here. For example, as shown in Figure \ref{fig:R0}c, if the first carbon atom occupy an energetically favorable hollow site, the next one closed to it will occupy a top site, which leads to a $E_f$ of 0.83 eV. Similar situation happens for armchair edges, carbon monomer attachment at some sites are energetically favorable, while it is not preferred at other sites. For an energetically favorable monomer attachment on the zigzag (armchair) edge, the energy barrier is 0.99 (1.15) eV. Since it is smaller than that of two carbon monomer combination, carbon monomers should have the capability of attaching to graphene edges at the growth temperature.

To make a prediction about the kinetics of R30 growth, attachment of carbon species with different sizes should be studied systematically. As shown in Figure \ref{fig:R30}, all clusters can be attached exothermically at some edge sites. For clusters larger than C$_2$, these attachment configurations already include one or more top sites occupied. Therefore, if C$_3$ is available, graphene can grow over difficult top sites. The possibility of growing over top site with even smaller C$_2$ is checked in Figure \ref{fig:R0}d, which suggests that C$_2$ attachment at top site is still energetically favorable, with a negative $E_f$ equal to -0.26 eV. Since the C$_2$ flux is much larger than those of larger clusters, the C$_2$ determined R30 phase growth will be much faster than that of R0 determined by larger clusters, and we can predict that the R30 phase growth will be more or less quadratic instead of quintic for the R0 phase.

\section{conclusion}

Atomistic details of graphene growth on Ir(111) surface have been revealed by first principles calculations. As building blocks of graphene, small carbon clusters are considered first. Arching chain structures are energetically favorable on terrace, and they are also of relatively high mobilities. Magic compact structures are identified at step edges, which are relevant to the nucleation process. Graphene nanoribbons are used as models of growth front. For the major R0 orientation, the growth is found to be quite inhomogeneous due to lattice mismatch. At some parts of the graphene edge, atomic carbon attachment is thermodynamically prohibited, and clusters (size around 5) attachment is required. The experimentally observed nonlinear relationship between growth rate and carbon monomer concentration is determined by such slow cluster attachment steps. A delicate kMC model is constructed to directly reproduce the experimental growth kinetics. As another result of lattice mismatch, graphene growth on Ir(111) is sensitive to the graphene orientation. The rotated R30 phase has less inhomogeneity, which leads to a faster growth rate. A quadratic growth behavior is predicted for the R30 phase growth. Atomistic mechanisms revealed here are essential for growth condition optimization in order to control graphene orientation and edge morphology, and they may also apply to other heteroepitaxial systems with a significant lattice mismatch.

\textbf{Supporting Information.} Potential energy and geometry of more carbon species on Ir(111) and Ir(332), more reaction paths of carbon species diffusion and coalescence, cluster attachment for armchair edges, energy barriers for carbon atom and clusters attachment, and more kMC simulation details.

\textbf{Acknowledgements.} We are grateful to Prof. Zhenyu Zhang for helpful discussions. This work is partially supported by NSFC (21173202, 20933006, 21121003, and 91021004), by MOST (2011CB921404), by CAS(KJCX2-YW-W22), and by USTC-SCC, SCCAS, and Shanghai Supercomputer Centers.

\end{document}